\documentclass[review]{elsarticle}
\usepackage[letterpaper, total={7.25in,9.5in}]{geometry}
\usepackage{amssymb}
\usepackage[]{lineno}
\usepackage{graphicx}
\usepackage{dcolumn}
\usepackage{bm}
\usepackage{multirow}
\usepackage[dvipsnames]{xcolor}
\usepackage[colorlinks,citecolor=blue,urlcolor=blue,bookmarks=false,hypertexnames=true]{hyperref}
\journal{Nuclear Instruments and Methods A}
\begin{document}

\begin{frontmatter}
\title{In-flight production of an isomeric beam of $^{16}$N}
\author[inst1]{C.~R.~Hoffman\corref{cor1}}
\cortext[cor1]{Corresponding author: crhoffman@anl.gov}
\affiliation[inst1]{organization={Physics Division, Argonne National Laboratory},
            city={Lemont},
            state={IL},
                        postcode={60439}, 
            country={USA}}
\author[inst1]{T.~L.~Tang\fnref{fn1}}
\fntext[fn1]{Present address: Department of Physics, Florida State University, Tallahassee,   Florida, 32306,USA}
\author[inst1]{M.~Avila}
\author[inst2]{Y.~Ayyad\fnref{fn2}}
\fntext[fn2]{Present address: IGFAE, Instituto Galego de F\'isica de Altas Enerx\'ias, Universidade de Santiago de Compostela, E-15782 Santiago de Compostela, Spain}
\affiliation[inst2]{organization={National Superconducting Cyclotron Laboratory},
            addressline={Michigan State University}, 
            city={East Lansing},
            state={MI},
                        postcode={48824}, 
            country={USA}}
\author[inst6,inst7]{K.~W.~Brown}
\affiliation[inst6]{organization={Facility for Rare Isotopes Beams, Michigan State University},
            city={East Lansing},
            state={MI},
                        postcode={48824}, 
            country={USA}}
\affiliation[inst7]{organization={Department of Chemistry, Michigan State University},
            city={East Lansing},
            state={MI},
                        postcode={48824}, 
            country={USA}}
\author[inst1,inst2]{J.~Chen}
\author[inst3]{K.~A.~Chipps}
\affiliation[inst3]{organization={Oak Ridge National Laboratory},
            state={TN},
                        postcode={37830}, 
            country={USA}}
\author[inst1]{H.~Jayatissa}
\author[inst1]{B.~P.~Kay}
\author[inst1]{C. M\"uller-Gatermann}
\author[inst4]{H.~J.~Ong}
\affiliation[inst4]{organization={Research Center of Nuclear Physics},
            addressline={10-1 Mihogaoka}, 
            city={Ibaraki, Osaka},
                        postcode={567-0047}, 
            country={Japan}}
\author[inst1]{J.~Song} %
\author[inst5,inst1]{G.~L.~Wilson}
\affiliation[inst5]{organization={Department of Physics and Astronomy},
            addressline={Louisiana State University}, 
            city={Baton Rouge},
            state={LA},
                        postcode={70803}, 
            country={USA}}

\begin{abstract}
An in-flight beam of $^{16}$N was produced via the single-neutron adding ($d$,$p$) reaction in inverse kinematics at the recently upgraded Argonne Tandem Linear Accelerator System (ATLAS) in-flight system. 
The amount of the $^{16}$N beam which resided in its excited 0.120-MeV $J^{\pi}=0^-$ isomeric state (T$_{1/2}\approx5$~$\mu$s) was determined to be 40(5)\% at a reaction energy of 7.9(3)~MeV/$u$, and 24(2)\% at a reaction energy of 13.2(2)~MeV/$u$. 
 The isomer measurements took place at an experimental station $\approx30$~m downstream of the production target and utilized an Al beam-stopping foil and a HPGe Clover detector. Composite $^{16}$N beam rate determinations were made at the experimental station and the focal plane of the Argonne in-flight radioactive ion-beam separator (RAISOR) with Si $\Delta$E-E telescopes. A Distorted Wave Born Approximation (DWBA) approach was coupled with the known spectroscopic information on $^{16}$N in order to estimate the relative $^{16}$N isomer yields and composite $^{16}$N beam rates. 
In addition to the observed reaction-energy dependence of the isomer fraction, a large sensitivity to the angular acceptance of the recoils was also observed. 
\end{abstract}

\end{frontmatter}


\section{\label{sec:intro} Introduction \& Background}

Radioactive beams containing substantial amounts of isomeric content provide a unique avenue for reaction and spectroscopic studies of nuclei.
A few applications of this approach are detailed in Refs.~\cite{ref:Watanabe2004,ref:Stefanescu2007,ref:Almaraz2017,ref:Santiago2018,ref:Asher2018,ref:Kahl2017,ref:Bowry2013,ref:Chipps2018,ref:Longfellow2020,ref:Tang2022,ref:Jones2022}.
Reaction methods such as direct single- or multi-nucleon transfer at energies up to $\sim$20~MeV/$u$ have shown promise in the production of isomeric beams.
This is due, in part, to the selective nature and prolific cross sections of these reactions, tools that have been utilized in nuclear science research for a number of years.
In addition, changes in the reaction energy or target species influence the relative yields to the final states of the secondary beam of interest.
These variables provide a measure of flexibility in the population of either the isomer or ground states as they typically have different underlying nuclear structure properties.

In the present work, we explored the production of the $^{16}$N nucleus which has an isomeric (meta-stable) state at an excitation energy of  E$_x=0.120$~MeV above the ground state~\cite{ref:Til93}. For simplicity, throughout the manuscript the isomeric state will be referred to as $^{16}$N$^{m}$, while the ground state will be labeled as $^{16}$N$^{g}$. The $^{16}$N$^{m}$ state has $J^{\pi}=0^-$, and the $^{16}$N$^{g}$ state $J^{\pi}=2^-$; the lifetimes of the two states are T$_{1/2}=5.25(6)~\mu$s and 7.13(2)~s, respectively~\cite{ref:Til93}. The isomeric level is formed due to the small binding-energy difference between it and the ground state in conjunction with a relatively large difference in $J$,  $\Delta J= 2$. The isomeric state decays with a $>99.99$\% branch to the ground state via a 120-keV $\gamma$-ray transition having an electric quadrupole ($E2$) multipolarity~\cite{ref:Til93,ref:Palffy1975,ref:Gagliardi1983,ref:Champagne1988}. There are two other known bound excited states in $^{16}$N at E$_x=0.298$~MeV ($J^{\pi}=3^-$) and 0.397~MeV ($J^{\pi}=1^-$), both with at least order-of-magnitude shorter lifetimes (T$_{1/2}\lesssim 100$~ps)~\cite{ref:Til93,ref:Kozub1983}. The 3$^-$ level decays via a $\gamma$-ray transition directly to the ground state with a 100\% branch. The excited 1$^-$ level feeds both the $^{16}$N$^{m}$ and $^{16}$N$^{g}$ states with branching ratios of 73.4(16)\% and 26.6(6)\%~\cite{ref:Til93}, respectively.

\section{\label{sec:meth} Methods \& Approach}

The application of transfer reactions in the production of in-flight beams has resulted in great success at accelerator facilities worldwide and the technique has been a component of the Argonne Tandem Linear Accelerator Systems (ATLAS) facility, located at Argonne National Laboratory, since the late 1990's~\cite{ref:Harss2000}.
Recently, an upgrade to the in-flight beam production capabilities of the ATLAS facility (the ATLAS in-flight system) was completed. The layout of the new ATLAS in-flight system and details into its characteristics are given in Section~\ref{sec:exp}.
The in-flight production reaction for $^{16}$N$^{m,g}$ was the well-studied single-neutron adding reaction of a $^{15}$N beam on a deuterium target in inverse kinematics, $^{15}$N($d$,$p$)$^{16}$N.
The $^{15}$N($d$,$p$) reaction has been carried out at various bombarding energies with relevance to nuclear structure and nuclear astrophysics exploration~\cite{ref:Warburton1957,ref:Bohne1972,ref:Bardayan2008}.

\begin{figure}[htb]
	\centering
    \includegraphics[width=0.5\textwidth]{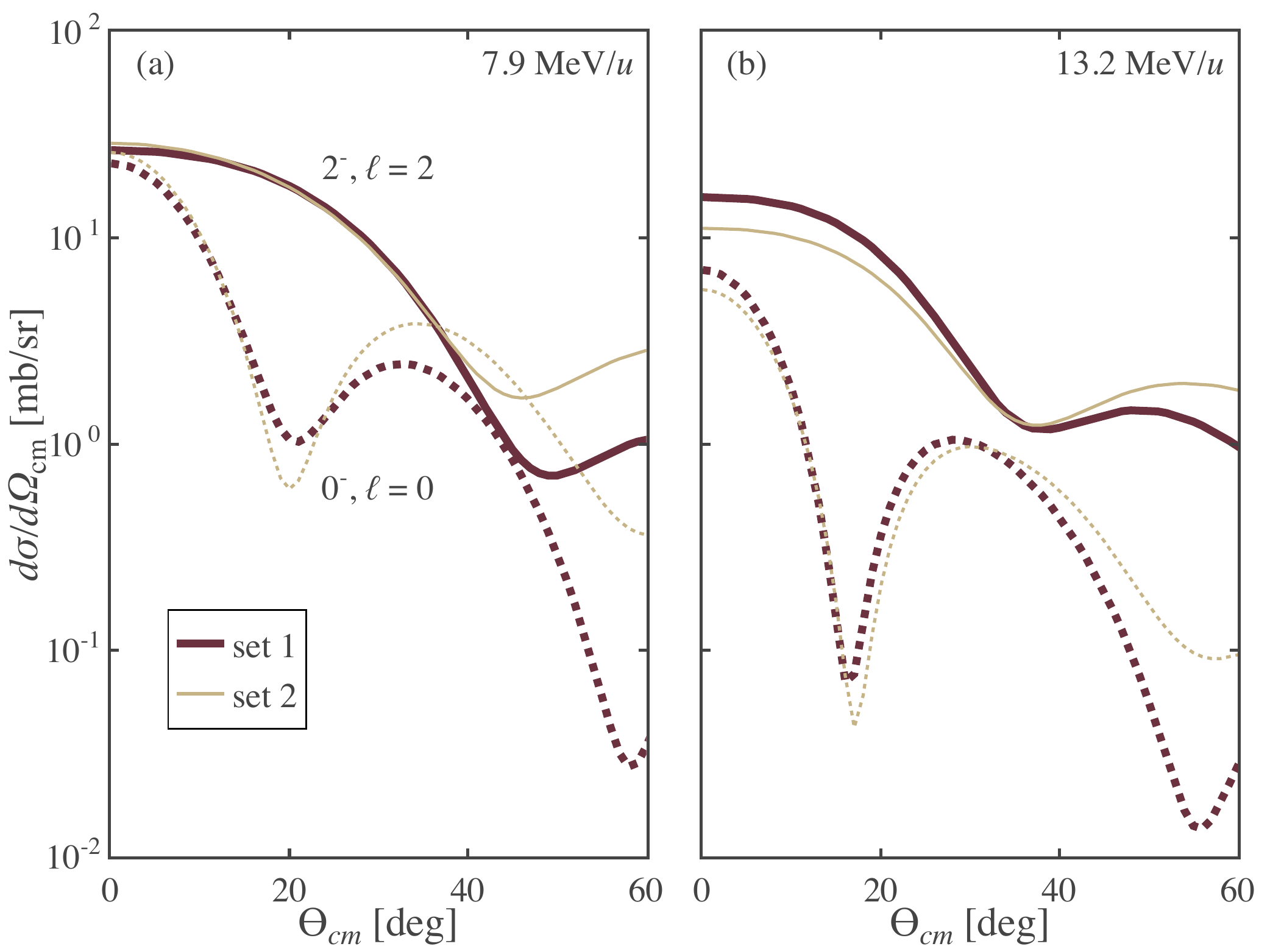}
	\caption{Calculated angular distributions (differential cross sections) $d\sigma/d\Omega_{cm}$ [mb/sr] for the $^{15}$N($d$,$p$)$^{16}$N reaction from the DWBA approach at reaction energies of (a) 7.9~MeV/$u$ and (b) 13.2~MeV/$u$. The solid (dashed) lines represent the neutron transfer from the $^{15}$N 1/2$^-$ ground state into the 2$^-$ (0$^-$) final bound state in $^{16}$N. The chosen DWBA optical model parameter sets 1~\cite{ref:Perey1963d,ref:Perey1963p,ref:Perey1976} and 2~\cite{ref:Becchetti1969,ref:An2006} are represented by the thick garnet (dark) and thin gold (light) lines, respectively.}
	\label{fig:ad}
\end{figure}

\begin{figure*}[htb]
	\centering
    \includegraphics[width=0.96\textwidth]{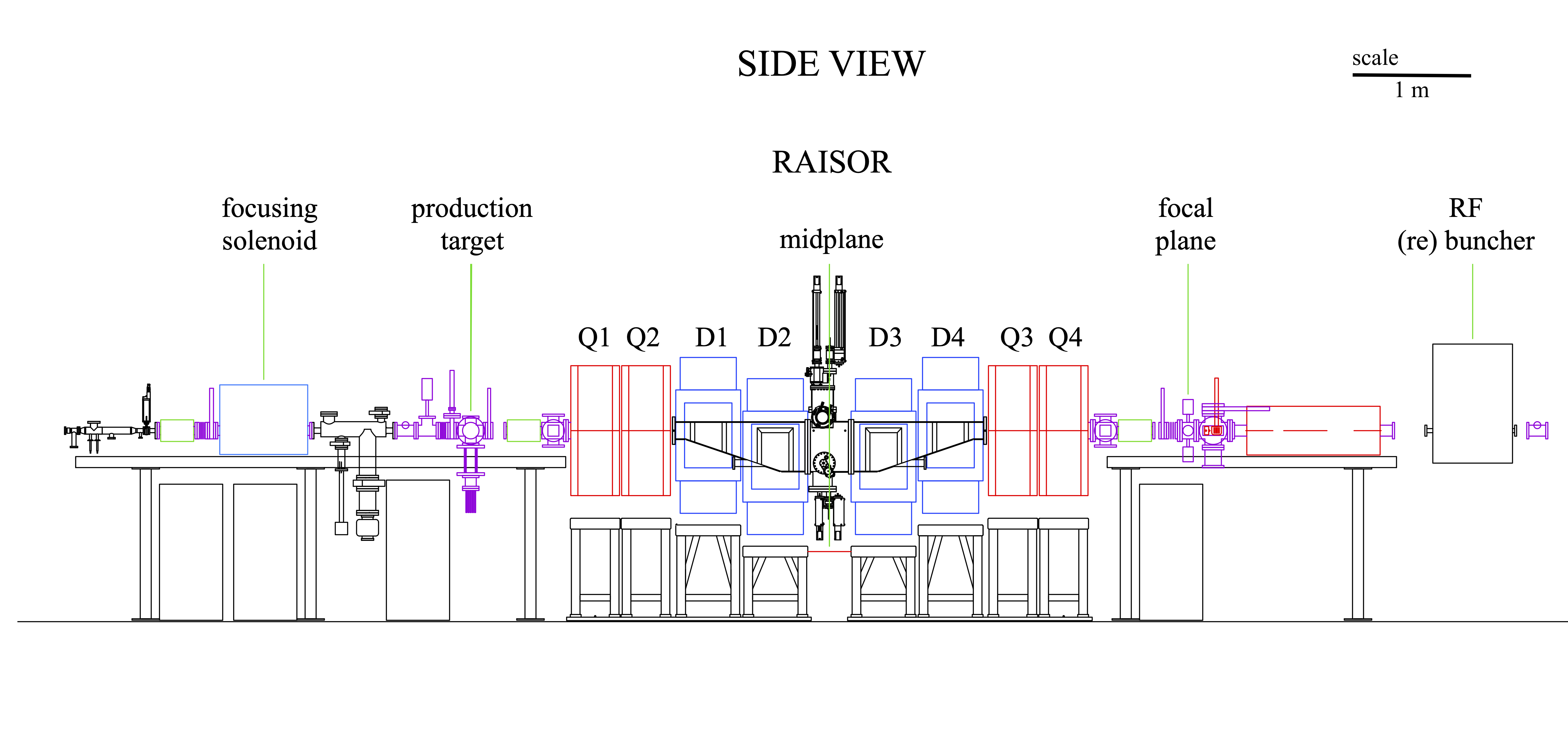}
	\caption{A partial (side) view of the layout of the upgraded ATLAS in-flight system. The beam direction is from left-to-right. The magnetic components of RAISOR, including the quadrupoles (Q\#) and dipoles (D\#), have been labeled. Not shown in the figure are the locations of the RF Sweeper and second RF (re)bunching cavity which are situated 21~m and 23~m downstream of the production target, respectively.}
	\label{fig:layout}
\end{figure*}

The two $^{16}$N bound states with largest $J$ (2$^-$, 0.000~MeV and 3$^-$, 0.298~MeV) are both directly populated through the transfer of a single neutron with an orbital angular momentum, $\ell$, equal to $2\hbar$ onto the $^{15}$N $J^{\pi}=1/2^-$ ground state~\cite{ref:Warburton1957,ref:Bohne1972,ref:Bardayan2008}. The two remaining states at 0.120~MeV (the 0$^-$ isomer) and 0.397~MeV (1$^-$), are both directly populated via an $\ell=0\hbar$ neutron transfer in the same reaction (the $\hbar$ will be omitted from this point forward for brevity). Differing $\ell$ transfer values give different population yields as a function of center-of-mass angle $\Theta_{cm}$, namely their differential cross sections, $d\sigma/d\Omega_{cm}$ [mb/sr] (Fig.~\ref{fig:ad}).
The relation of $\Theta_{cm}$ to the laboratory angle of a $^{16}$N recoil, $\Theta_{lab}$, for these inverse kinematic reactions is $\Theta_{cm}\approx10^{\circ}(45^{\circ})\rightarrow\Theta_{lab}\approx1^{\circ}(4^{\circ})$, and remains similar over energies of interest (3 -- 16~MeV/$u$).
Therefore, a direct relation exists between the acceptance in $\Theta_{lab}$ and varying populations of the states in $^{16}$N and ultimately leading to changes in the $^{16}$N$^{m}$-to-$^{16}$N$^{g}$ ratio.

\subsection{the Distorted Wave Born Approximation approach}
There are standard tools and techniques available to calculate the relative integrated differential cross sections, or simply cross sections, $\sigma$ [mb], to the known bound states in $^{16}$N via the ($d$,$p$) reaction.
The Distorted Wave Born Approximation (DWBA) approach has been adopted in the present work due to its reliable calculation of single-neutron transfer reaction cross sections in the energy range of interest.
The DWBA approach also calculates an angular distribution corresponding to the unique differential cross sections observed for each neutron $\ell$ transfer. 
The $\sigma$ from the DWBA approach are defined by the integration of the differential cross sections over $\Theta_{cm}$ from 0 up through a maximum angle, $\Theta_{cm}^{max}$. The cross sections are combined with known spectroscopic information about $^{16}$N to provide estimates of the $^{16}$N$^m$ fraction in the beam.
The required properties of $^{16}$N include the relative spectroscopic overlaps ($S_{\ell}$), which have been extracted from the $^{15}$N($d$,$p$) reaction data~\cite{ref:Warburton1957,ref:Bohne1972,ref:Bardayan2008}, as well as the measured relative $\gamma$-ray decay branches from the excited 1$^-$ and 3$^-$ levels~\cite{ref:Til93}.

\begin{figure*}[htb]
	\centering
    \includegraphics[width=0.48\textwidth]{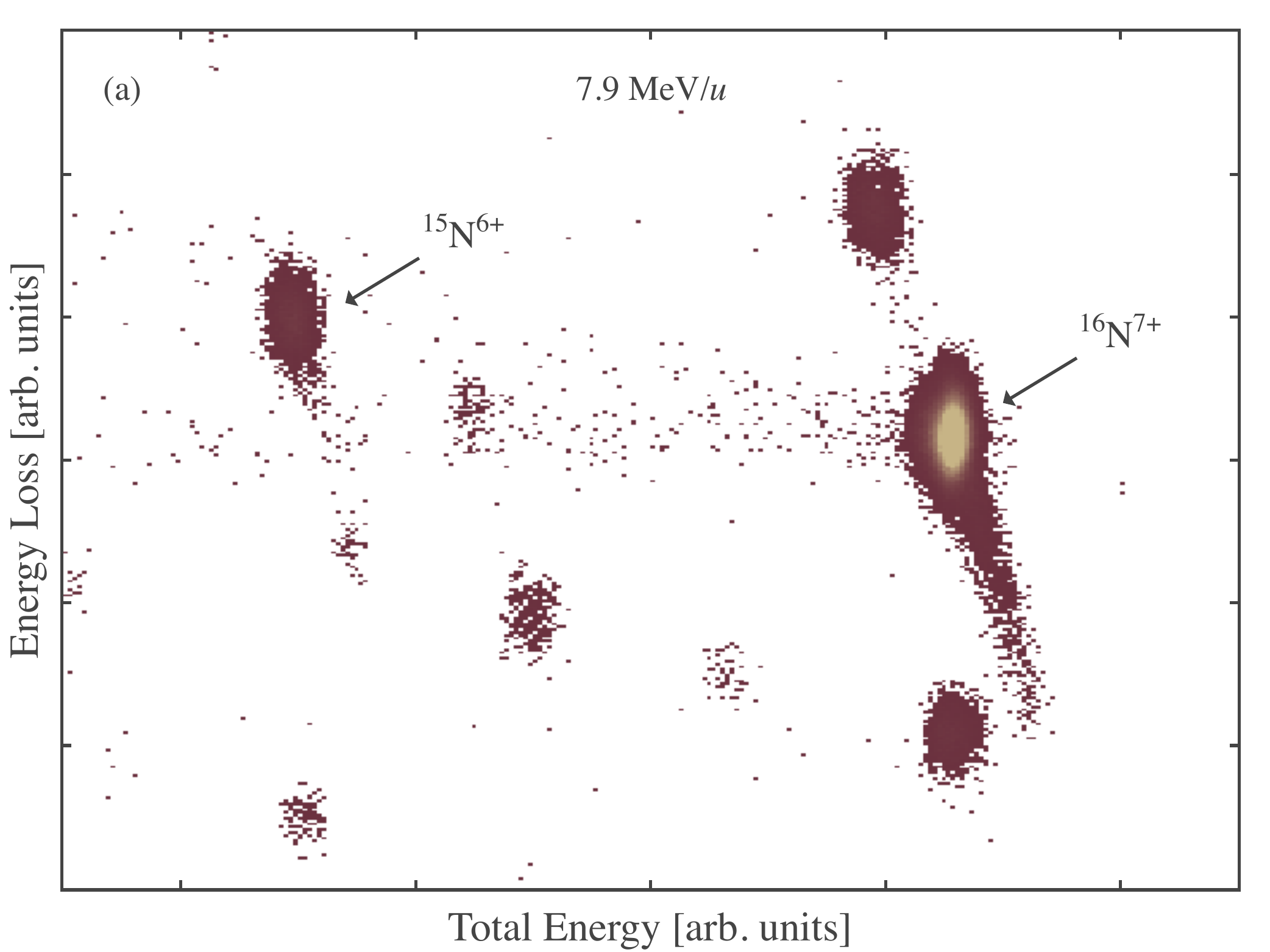}
    \includegraphics[width=0.48\textwidth]{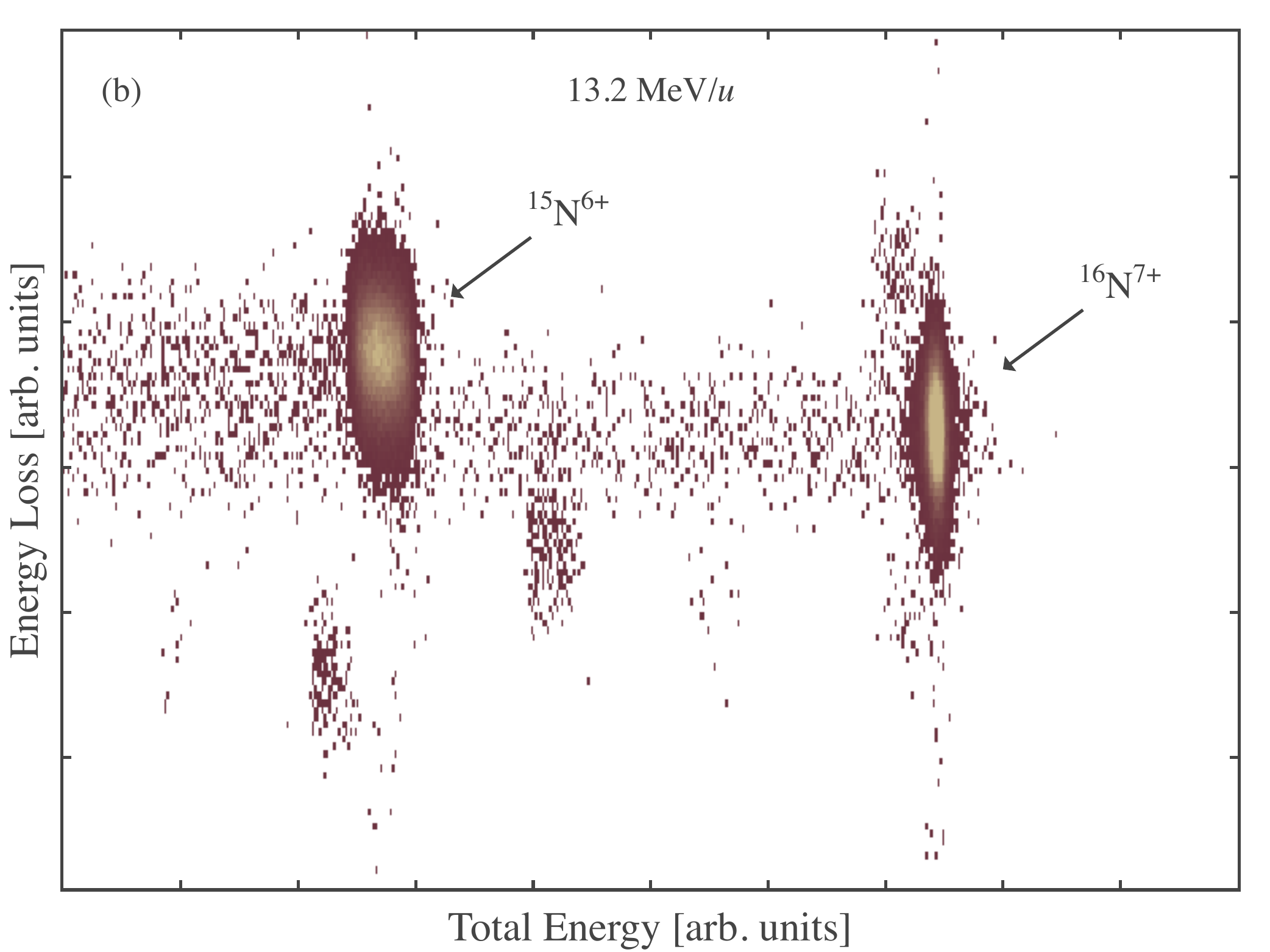}
	\caption{Identification of the $^{16}$N secondary beam at the experimental station by the Si $\Delta$E-E telescope for reaction energies of 7.9(3)~MeV/$u$ (a) and 13.2(2)~MeV/$u$ (b). The largest beam contaminant, $^{15}$N$^{6+}$, is also labeled in each plot.}
	\label{fig:pid}
\end{figure*}

The DWBA calculations were done using the software code PTOLEMY~\cite{ref:Mac78}. 
Fig.~\ref{fig:ad} shows the calculated $d\sigma/d\Omega_{cm}$ for the 2$^-$ and 0$^-$ states populated in the $^{15}$N($d$,$p$)$^{16}$N reaction. The angular distributions for the $3^-$ and $1^-$ states are similar for their respective $\ell$ values.
A number of optical model parameter sets were investigated for the distorting potentials of the partial waves. 
Global parameter sets were preferred due to their inherent energy dependencies as summarized in Ref.~\cite{ref:Perey1976}. 
The two sets shown in Fig.~\ref{fig:ad} both described the trends of the $\ell=2$ and $0$ data collected in Refs.~\cite{ref:Warburton1957,ref:Bohne1972,ref:Bardayan2008}.
Set 1 (thick garnet lines) consisted of the proton parameters of Ref.~\cite{ref:Perey1963p} and the deuteron parameters of Ref.~\cite{ref:Perey1963d}. 
Set 2 (thin gold lines) consisted of the proton parameters of Ref.~\cite{ref:Becchetti1969} and the deuteron parameters of Ref.~\cite{ref:An2006}.
A Woods-Saxon potential was used to describe the final bound states in $^{16}$N with a potential radius given by r=r$_0A^{1/3}$, where r$_0= 1.25$~fm, and a potential diffuseness of a$_0 = 0.65$~fm.
The potential depth was varied independently to reproduce the binding energies of each $^{16}$N final state.
The deuteron bound-state was described by the Argonne V$_{18}$ potential~\cite{ref:Wiringa1995}.
Considering that an estimation of the isomer content only relied on the relative $\ell=2$ to $\ell=0$ cross sections, variance between different optical model parameters sets was reduced.
Similarly, sensitivities to bound state parameters were not a significant source of uncertainty in the isomer fraction estimates.

\subsection{the isomer fraction determination method}
The empirical determination of the isomer content in the $^{16}$N beam was made by comparing the number of isomer state decays via the 120-keV $\gamma$-ray transition~\cite{ref:Gagliardi1983} to the number of total ground state decays via known $\beta$-delayed $\gamma$-ray transitions~\cite{ref:Alburger1958,ref:Alburger1959}. See, for example, the first figures in both Ref~\cite{ref:Alburger1958} and~\cite{ref:Champagne1988}, or Figure 2 in Ref.~\cite{ref:Gagliardi1983}, for a depiction of the isomer and ground-state decay schemes.
 From the $^{16}$N ground state $\beta$-decay, the 1755-keV and 2742-keV transitions de-exciting the 8.872-MeV level in $^{16}$O~\cite{ref:Wilkinson1956} with decay branches of 0.121(10)\% and 0.82(6)\%, respectively, were used. While more prevalent transitions do appear at $\sim$6 and $\sim$7~MeV in the $\beta$-delayed decay scheme, they were not used because determination of their energy-dependent peak efficiencies was unreliable (additional details in Section~\ref{sec:exp}).   
The number of $^{16}$N$^{m}$ nuclei that were implanted over a fixed time duration was extracted from the measured area of 120-keV peak in the $\gamma$-ray spectrum, $N_m$, after being combined with the corresponding energy-dependent peak efficiency and decay branch information.
Similarly, the total number of $^{16}$N$^{m+g}$ beam ions that were implanted over the same time duration was found from the measured areas of the 1755-keV and 2742-keV $\gamma$-ray peaks, $N$, their known decay branches, and their energy-dependent peak efficiencies. 
The isomer fraction and beam composition, $^{16}$N$^{m}/^{16}$N, was revealed through the ratio of the number of $^{16}$N$^{m}$ nuclei to the total number of $^{16}$N$^{m+g}$ ions implanted. The length of the counting periods were long ($\approx$hours) relative to the isomer and ground state half-lives (T$_{1/2}\approx5~\mu$s, and $\approx7$~s), and the ground state of the daughter nucleus, $^{16}$O, is stable. This ensured that explicit handling of the build-in and decay-out structures or other half-lives was not necessary.

\section{\label{sec:exp} Experiment \& Results}
\subsection{the ATLAS in-flight system}
The selection and delivery of the $^{16}$N secondary beam was carried out using the recently upgraded ATLAS in-flight system (Fig.~\ref{fig:layout}).
The centerpiece of the upgrade was the implementation of the Argonne in-flight radioactive ion-beam separator (RAISOR)~\cite{ref:Mustapha2014}.
RAISOR was developed to provide momentum selection of a secondary ion beam via its corresponding magnetic rigidity (B$\rho$) while also providing the means to suppress and stop any high-intensity unreacted primary beam. In addition, RAISOR was designed as a magnetic achromat from target to focal plane, in order to facilitate the delivery of secondary beams from its focal plane to any of a number of experimental stations via tens-of-meters of beam transport lines. 
The symmetric layout of the RAISOR magnetic chicane consists of four magnetic dipoles with bookend pairs of quadrupole doublets as shown in Fig.~\ref{fig:layout}. 
The maximum field strength of the RAISOR magnets are 1.75~T for the dipole magnets and 1~T at the field tips of the quadrupole magnets.
The energy dispersion at the midplane is 1.3~mm/\%, with a magnification factor of 0.6, for a midplane crossing distance of $30$~cm off from the beam axis.
The ion flight-path length through the device is $\sim$7~m for the same 30~cm beam displacement (the linear on-axis distance from the production target to the RAISOR focal plane is 6.6~m).
The angular acceptance of the device is $75$~mrad as defined by the 15 cm inner beam-line chamber diameter at the location of the first quadrupole. 
The energy acceptance of RAISOR can reach $\Delta$E/E~=~10\% due to the broadband design of the device. Adjustable water-cooled slits have been placed at the midplane and are used in the B$\rho$ selection and energy acceptance selection of the beam of interest. The midplane slits also define the controlled stopping location of the high-intensity unreacted primary beam.

\begin{table*}[htb]
\caption{The energy-dependent efficiencies ($\epsilon$) extracted for the 1755-keV and 2742-keV $\gamma$ rays relative to the 120-keV transition for each HPGe Clover crystal and for both beam energies are listed along with the measured areas ($N_m$ and $N$) and extracted $^{16}$N$^{m}$/$^{16}$N isomer fraction values. The $N_m$ and $N$ values include statistical and fit-related uncertainties. The $^{16}$N$^{m}$/$^{16}$N values include additional systematic uncertainties of 3\% and 5\% from the $\epsilon$, and 8\% and 7\% on the decay branches, from the 1755-keV and 2742-keV $\gamma$-ray transitions, respectively. In addition, the total $^{16}$N beam rates determined at the experimental station and at the RAISOR focal plane (at 13.2~MeV/$u$ only) are listed with errors.}\label{tab:rates}
\centering
	\begin{tabular}{ccccccccccc}
	\hline \hline
	Energy & $^{16}$N rate [x10$^3$] & \multirow{2}{*}{crystal} & \multicolumn{2}{c}{$\epsilon$ [\%]$^{a}$} & $N_m$ [x10$^3$] & \multicolumn{2}{c}{$N$ [x10$^3$]} & \multicolumn{2}{c}{$^{16}$N$^{m}$/$^{16}$N}  \\
	 MeV/$u$ & [pps/pnA] &        &         1755 & 2742 & 120 & 1755 & 2742 & 120/1755 & 120/2742  \\
	            	\hline
	 \multirow{3}{*}{7.9(3)} & \multirow{3}{*}{6(2)}& 1     & 20.7  & 17.1 &   663(1) & 0.408(0.070) & 2.13(0.07)  & 0.406(0.077) &  0.436(0.040) \\
	                         && 2     & 19.7  & 15.1 &   687(1) & 0.392(0.078) & 2.30(0.07)  & 0.417(0.091) &  0.371(0.034) \\
	                         && 3     & 20.3  & 16.0 &   574(1) & 0.310(0.061) & 2.12(0.07)  & 0.455(0.099) &  0.355(0.033) \\
	     \\
	\multirow{3}{*}{13.2(2)} & \multirow{2}{*}{2(1)}& 1     & 20.7  & 17.1 & 2720(2)  &  3.01(0.17) &  14.5(0.2) & 0.226(0.023) & 0.262(0.023) \\
	                         & \multirow{2}{*}{27.5(2.5)$^b$}& 2     & 19.7  & 15.1 & 3360(2)  &  3.52(0.29) &  16.8(0.2) & 0.227(0.027) & 0.248(0.022) \\
	                         && 3     & 20.3  & 16.0 & 2683(2)  &  3.29(0.23) &  14.8(0.2) & 0.201(0.022) & 0.238(0.021) \\
	                         \hline\hline
	\end{tabular}
	\begin{flushleft}
	  $^{a}$Relative to the 120-keV $\epsilon\equiv100$\%.\\
	  $^b$Measured at the RAISOR focal plane.
	\end{flushleft}
\end{table*}

The target area of the ATLAS in-flight system, located on the upstream side of RAISOR (Fig.~\ref{fig:layout}) includes a 6-T solenoid for focusing the primary beam onto the production target and the capability to use self-supported foils or a cryogenically-cooled gas cell~\cite{ref:Rehm2011} as production targets. In the present work, the gas cell was cooled to $90$~K, filled to 1400~mbar with D$_2$ gas, and confined by HAVAR entrance and exit windows of thickness 1.9~mg/cm$^2$. Downstream of the RAISOR focal plane and located throughout the transport beam line are two superconducting Radio-Frequency (RF) resonators and an RF Sweeper~\cite{ref:Pardo2015}. The RF resonators provide control of the longitudinal phase ellipse of the secondary beam in order to minimize either the energy-spread or time-spread of the beam through (re-)bunching. The RF Sweeper facilitates improved beam purity through velocity selection and is based on an application of a transverse electric field oscillating at half the primary beam pulse frequency~\cite{ref:Pardo2015}. Reasonable beam purity ($>50$\%) was achieved for the $^{16}$N secondary beam from the RAISOR midplane slit settings alone (Fig.~\ref{fig:pid}), and the RF resonators and RF Sweeper elements were not used in the present work.

\subsection{the $^{15}$N primary beam}
The primary beam of $^{15}$N was delivered by ATLAS at two different energies, 8.1~MeV/$u$ and 13.5~MeV/$u$, and with available intensities up to 150~pnA.
The energy of the unreacted primary (degraded) beam was measured after the production target to be 7.4~MeV/$u$ and 12.7~MeV/$u$ for the two energies with a precision of $\sim$0.5\% via the ATLAS time-of-flight RF pick-up system.
Based on energy-loss calculations carried out with LISE++~\cite{ref:Tarasov2016}, the energies of the primary beams were estimated to range from 7.6 -- 8.1~MeV/$u$ and 13.0 -- 13.4~MeV/$u$ while traversing through the D$_2$ gas region of target.
Therefore, the calculated mid-target primary beam energies of 7.9(3)~MeV/$u$ and 13.2(2)~MeV/$u$ have been adopted as the average reaction energies corresponding to the two isomer fraction determinations.

\subsection{the $^{16}$N secondary beam}
The peak of the production yield for the fully-stripped $^{16}$N$^{7+}$ ion was determined through a systematic B$\rho$ scan of the RAISOR magnetic elements while monitoring of the composite $^{16}$N$^{7+}$ rate at the RAISOR focal plane via the Si $\Delta$E-E telescope.
The optimal rigidity was found to be $\Delta$B$\rho=+2.5$\% above the reference B$\rho$ from the measured $^{15}$N$^{7+}$ primary degraded beam. The same $\Delta$B$\rho$ value was found for each beam energy. The final beam settings had a momentum acceptance of $\Delta$P/P$\approx$2\% as defined by the RAISOR midplane slit spacing ($\sim 5$~mm). The product of the $\Delta$B$\rho$ scale factor ($\equiv1.025$) with the measured B$\rho$ of the primary degraded beam determined $^{16}$N$^{7+}$ secondary beam energies of $6.8$~MeV/$u$ and 11.8~MeV/$u$, with an error of a few percent in total energy on each. The $^{16}$N$^{7+}$ secondary beam was transported from the RAISOR focal plane to the experimental station by first applying the 1.025 scale factor to each magnetic element in the transport beam line. Then, some manual optimization of the transport beam-line elements was completed by the ATLAS operations staff. The minuscule change in the mass/momentum between the $^{16}$N$^{m}$ and $^{16}$N$^{g}$ beam components caused by the population of the 120-keV isomer state, would not impact their relative transport efficiency. However, as noted above and discussed below, the $^{16}$N$^{m}$/$^{16}$N ratio is sensitive to the angular acceptance, and/or the momentum acceptance through the reaction kinematics, because of the different $d\sigma/d\Omega_{cm}$ for $\ell=0$ versus $\ell=2$ neutron transfer at beam formation.

The $^{16}$N beam identification, composite rate determinations, and purity measurements were carried out using the Si $\Delta$E-E telescopes with thicknesses of $\Delta$E~=~25~$\mu$m and E~=~1000~$\mu$m.
One telescope was located at the focal plane of RAISOR, while a second telescope was placed at the experimental station, $\sim$30~m downstream of the in-flight production target.
$^{16}$N secondary beam identification plots taken at the experimental station are shown in Figs.~\ref{fig:pid}(a) and (b) for the two beam energies. 
The beam purity in both cases was better than $50$\% with the main beam contaminant being a low-energy component of the $^{15}$N$^{6+}$ degraded beam.
Some additional, weaker contaminants ($<1$\% of the total beam composition) were also identified, for instance, $^{16}$O$^{7+}$ populated via the $^{15}$N($d$,$n$) reaction.
The average rate of $^{16}$N detected at the experimental station was $4-8\times10^3$ and $1-3\times10^3$ particles-per-second per particle-nano-Ampere of primary beam current (pps/pnA) for the low- and high-energy beam settings, respectively.
The primary beam current was measured from the charge integration of a Faraday cup located upstream of the in-flight system production target.
A composite $^{16}$N rate of $25-30\times10^3$~pps/pnA for the high-energy setting only was measured at the RAISOR focal plane via $\Delta$E-E telescope. The main sources of uncertainty on the rate determinations were the measurement of and fluctuations in the primary beam current intensity.

\begin{figure}[t]
	\centering
    \includegraphics[width=0.48\textwidth]{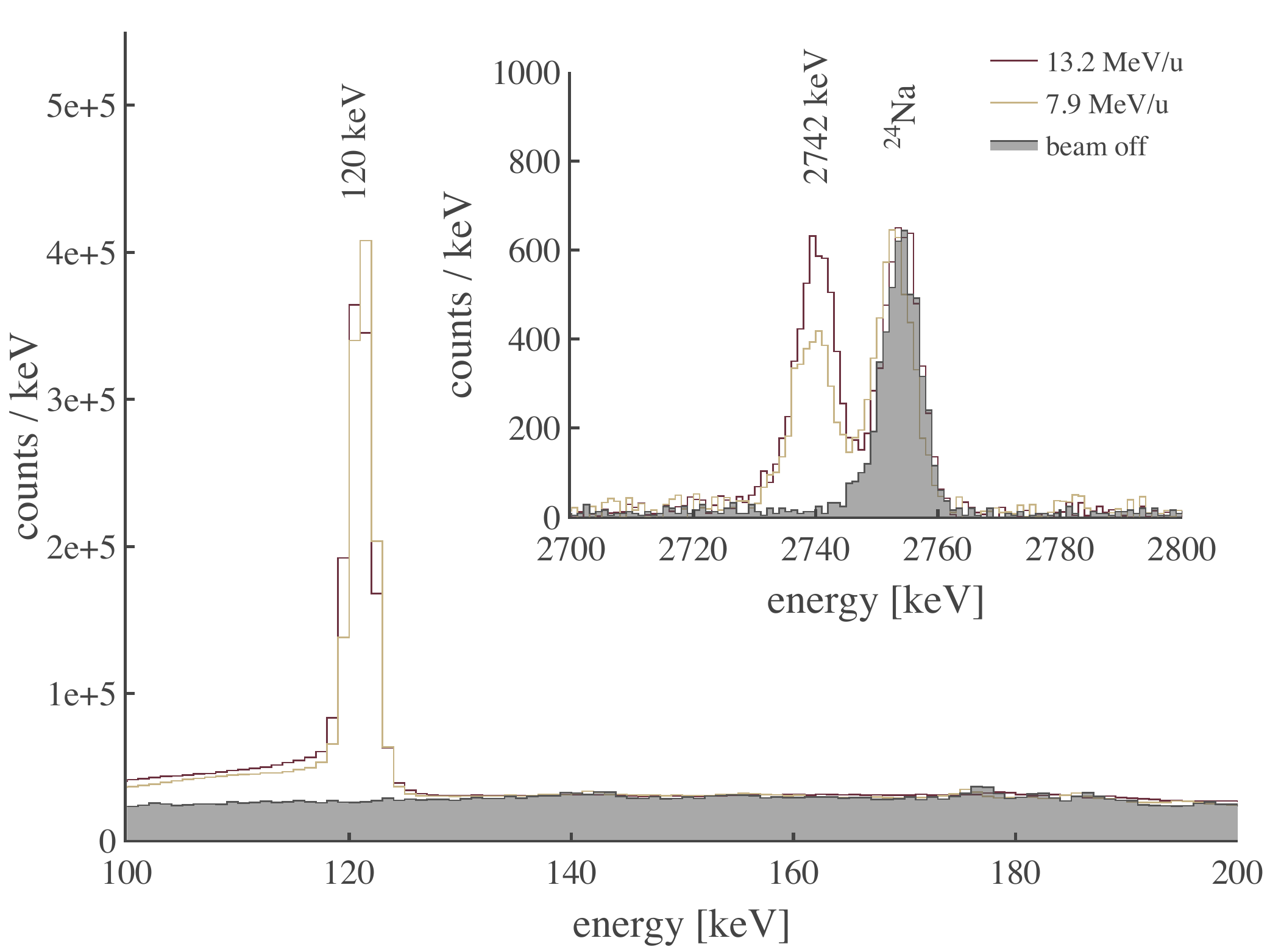}
	\caption{$\gamma$-ray spectra from one of the HPGe Clover crystals in the regions surrounding the 120-keV and 2742-keV (inset) transitions of interest. Data are included from the 13.2~MeV/$u$ (garnet lines) and 7.9~MeV/$u$ (gold lines) reaction energies, as well as, a background spectrum taken during a beam-off period (grey shaded area and lines). The two beam-on histograms have been normalized to the area of the 120-keV transition. The beam-off histograms were normalized independently to the background level around the 120-keV transition and to the area of the 2754-keV line from the $^{24}$Na $\beta$-decay contaminant peak in the inset.
	}
	\label{fig:gamma}
\end{figure}

\subsection{the gamma-ray data \& results}
Measurement of the $^{16}$N isomeric content physically took place at the same experimental station and beam-line location as the $\Delta$E-E Si telescope array.
The $\gamma$-ray detection setup consisted of an Al stopping/implant foil of thickness $\approx 500$~$\mu$m and a HPGe Clover detector consisting of three independently functioning HPGe crystals.
When in use, the Al foil was inserted in place of the $\Delta$E-E Si telescope and the foil thickness ensured all $^{16}$N beam particles came to a rest within its volume ($\lesssim350\mu$m).
The foil was arranged at a $45^{\circ}$ angle relative to the incoming beam.
The HPGe Clover detector measured $\gamma$-ray transitions following either the isomer decay or the ground state $\beta$ decay.
It was placed $\sim$13~cm from the Al stopping/implant foil and perpendicular to the beam direction.
Each of the three functioning crystals of the HPGe Clover detector was self-triggered and readout by a digital data acquisition system with a signal sampling rate of 100-MHz. The collection of the $\gamma$-ray data occurred over a continuous measuring period of $\approx12$~hrs while the $^{16}$N beam was being implanted at the reaction energy of 13.2~MeV/$u$ and with a primary beam intensity of 20~pnA.
Similarly, at the lower-energy beam setting an average primary beam intensity of 6~pnA was used over a continuous observation duration of $\approx8$~hrs. In both cases the approximate rate of $^{16}$N impinging on the implant foil was $\approx4\times10^4$~pps throughout the data collection period.

The area of each $\gamma$-ray transition of interest was extracted using a Gaussian line-shape fitting procedure which included a linear background component. The resulting areas ($N_m$ and $N$) are listed in Table~\ref{tab:rates} for the three $\gamma$-rays of interest, at both beam energy settings, and for each of the three HPGe Clover crystals. Two of the energy regions of interest are shown in Fig.~\ref{fig:gamma} for one of the HPGe Clover detector crystals. The 120-keV transition belonging to the $^{16}$N$^{m}$ 0$^-\rightarrow 2^-$ decay is labeled.
The inset also shows the 2742-keV transition in $^{16}$O coming from the $\beta$-delayed decay of $^{16}$N$^{g}$. All other prominent transitions expected in the decay of $^{16}$N$^g$ were also identified, including the 6129-keV and 7115-keV $\gamma$ rays~\cite{ref:Til93}. The 7.9~MeV/$u$ (gold line) and 13.2~MeV/$u$ (garnet line) histograms have been normalized to each other through the area of the 120-keV $\gamma$ ray. The beam-off data (grey shaded area and line) was collected after the high-energy beam setting and its histogram was normalized to the average background region for histogram containing the 120-keV transition. The beam-off spectrum was separately normalized to the area of the $^{24}$Na 2754-keV decay peak in the inset. In both cases, the transitions belonging to the ground-state and isomeric states of $^{16}$N do not appear in the beam-off spectra, as expected. As noted above, the contaminant peak belongs to the known 2754-keV transition in $^{24}$Mg following the $\beta$-decay of the $^{24}$Na ground state~\cite{ref:Firestone2007}.

In the processing of the $\gamma$-ray data, no add-back, energy-summing or coincidence procedures were applied to either the source data or the isomer data for the three crystals. Energy response calibrations and the energy-dependent full-energy peak efficiencies, $\epsilon$, for each crystal were extracted from standard $\gamma$-ray sources. The isotopes of $^{152}$Eu, $^{207}$Bi, $^{133}$Ba, and $^{137}$Cs provided energies up to 1770 keV. The sources were placed at the Al foil position within the beam-line chamber to account for the attenuation/scattering of $\gamma$ rays by the vacuum chamber material and other nearby materials. The spatial implant region of the $^{16}$N beam may cover a region spanning $\sim$5--8~mm on the Al stopping foil due to the in-flight beam spot size and drifts in the beam transport line. Hence, various positions of the source were explored and no sensitivity in the resulting energy dependent efficiency curve was observed. A model and simulation of the setup was developed in GEANT4~\cite{ref:Agostinelli2003}, and was used to validate the shape and magnitude of the energy-dependent efficiency curve.

Only the relative $\epsilon$ values were required between the 120-keV transition and either the 1755-keV or 2742-keV $\gamma$ rays for the $^{16}$N$^m$/$^{16}$N determination. Error estimates of 3\% and 5\% were adopted for these relative $\epsilon$ values (Table~\ref{tab:rates}). The absolute $\epsilon$ for the 120-keV transition was on the order of 2-3\%. An additional check of the relative $\epsilon$ curve was also made using two previously known transitions in $^{24}$Mg at 1368 keV and 2754 keV~\cite{ref:Firestone2007}. The transitions appear in the present work through the $\beta$-decay of the $^{24}$Na $J^{\pi}=4^{+}$ ground state (T$_{1/2}=15$~hr), which was likely created in the fusion evaporation reaction of the $^{16}$N beam on the Al stopping foil. The efficiency corrected transitions agree with their expected one-to-one branching ratio to within the uncertainties on the relative efficiencies.

\subsection{the $^{16}$N$^m$/$^{16}$N isomer fraction results}

The extracted values of $^{16}$N$^m$/$^{16}$N for each of the three crystals, both beam energies, and both the 1755-keV and 2742-keV transitions, are given in Table~\ref{tab:rates}. The $N_m$ and $N$ areas, $\epsilon$ percentages, and the experimental branching ratios required to derive the $^{16}$N$^m$/$^{16}$N values are also listed in the table or the caption. For all but the isomer ratio extracted from the 1755-keV line at 7.9~MeV/$u$ which was dominated by statistics, the dominant source of error on $^{16}$N$^m$/$^{16}$N is the uncertainty on the decay branches of the 1755-keV and 2742-keV transitions.
The weighted averages of the isomer fraction from the 120-keV/1755-keV data were 43(8)\% and 22(2)\%, for the lower and higher reaction energies, and similarly, 39(4)\% and 25(2)\% for the 120-keV/2742-keV data.
The isomer fraction results are plotted as a function of reaction energy -- the calculated mid-target beam energy -- in Fig.~\ref{fig:results}. 

\section{\label{sec:dis} Discussion}
The change in the measured isomer content as a function of reaction energy (Fig.~\ref{fig:results}) highlights one of the degrees of freedom which determines the $^{16}$N beam composition. 
The increase in isomer content towards lower reaction energy is understood through straight-forward arguments surrounding the physics governing single-neutron transfer reactions.
Namely, it has been empirically established that single-nucleon transfer reactions exhibit integrated yields which peak at different reaction energies dependent upon the $\ell$ value of the transferred particle.
This is due to a momentum matching condition between the beam-target system and the orbital angular momentum $\ell$ being transferred, see Refs.~\cite{ref:Sch12,ref:Sch13}, for example.
In the present work, improved momentum-matching occurs for the $\ell=0$ neutron transfer as the reaction energy is lowered, and vice-versa for an $\ell=2$ neutron transfer. 
The momentum-matching impact can be viewed in Fig.~\ref{fig:ad} where there is relative increase in the $\ell=0$ differential cross sections between those at 13.2~MeV/$u$ in (b) versus those at 7.9(3)~MeV/$u$ in (a). 
At lower reaction energies the effect is an increase in the $^{16}$N$^m$/$^{16}$N fraction, considering that $\approx80$\% of the $\ell=0$ cross section feeds into the $0^-$ isomeric state.

\subsection{\label{subsec:dwba}the estimation of the isomer fraction}
\begin{figure}[t]
	\centering
    \includegraphics[width=0.48\textwidth]{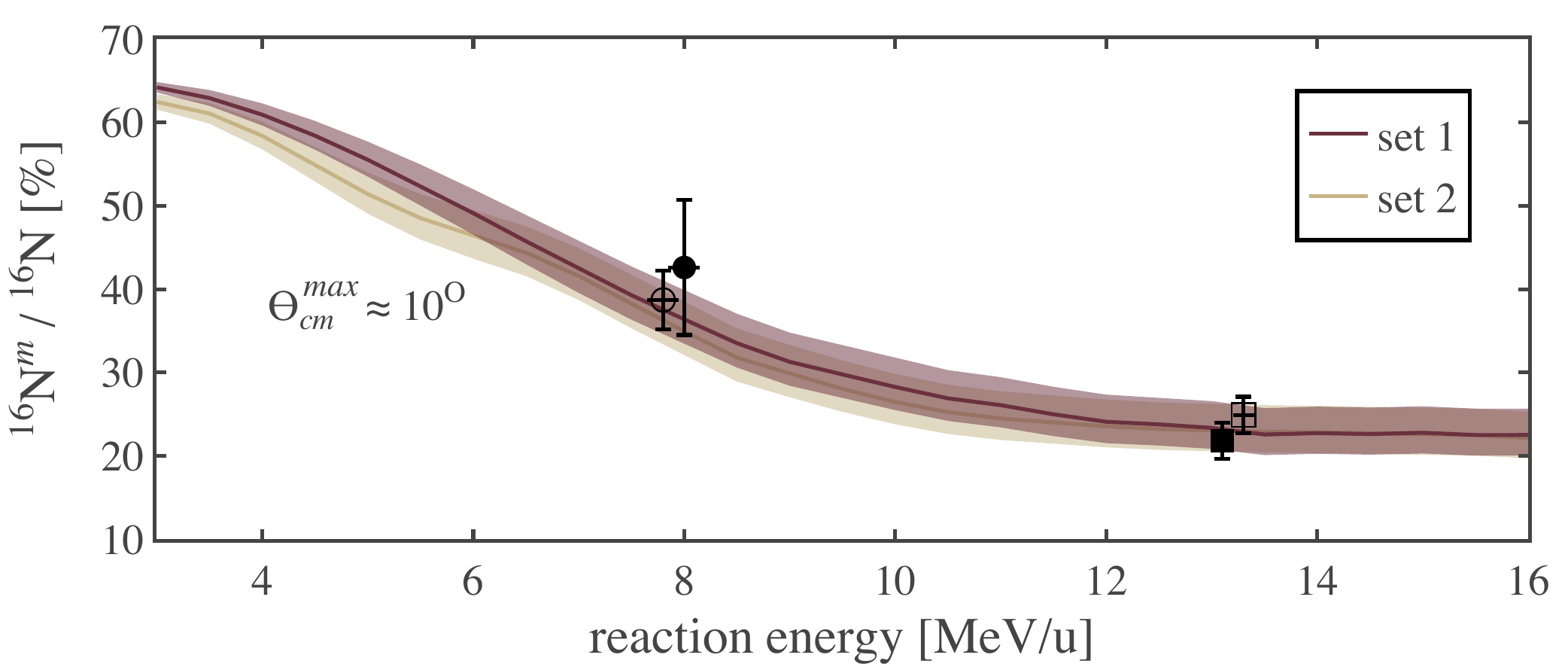}
	\caption{The $^{16}$N$^{m}$/$^{16}$N isomer fraction, given in percent, as a function of the $^{15}$N$(d$,$p$) reaction energy (MeV/$u$). The data is represented by the black points for both the low (squares) and high (circles) reaction energies as well as for both the 1755-keV (open) and 2742-keV (filled) $\gamma$-ray transitions. Uncertainties on the isomer percent and the reaction energy are reflected in the width of the points and the error bars. The calculated isomer percent from the DWBA approach is shown for $\Theta_{cm}^{max}\approx10^{\circ}$. The solid lines correspond to $S_{2}/S_{0}=0.85$, and the lower-to-upper limits of the shaded areas correspond to $S_{2}/S_{0}=0.7-1.0$, respectively. The two different DWBA optical model parameter sets, set 1~\cite{ref:Perey1963d,ref:Perey1963p,ref:Perey1976} and set 2~\cite{ref:Becchetti1969,ref:An2006}, are represented by the garnet (dark) and gold (light) colors, respectively.
	}
	\label{fig:results}
\end{figure}

\begin{figure}[t]
	\centering
    \includegraphics[width=0.48\textwidth]{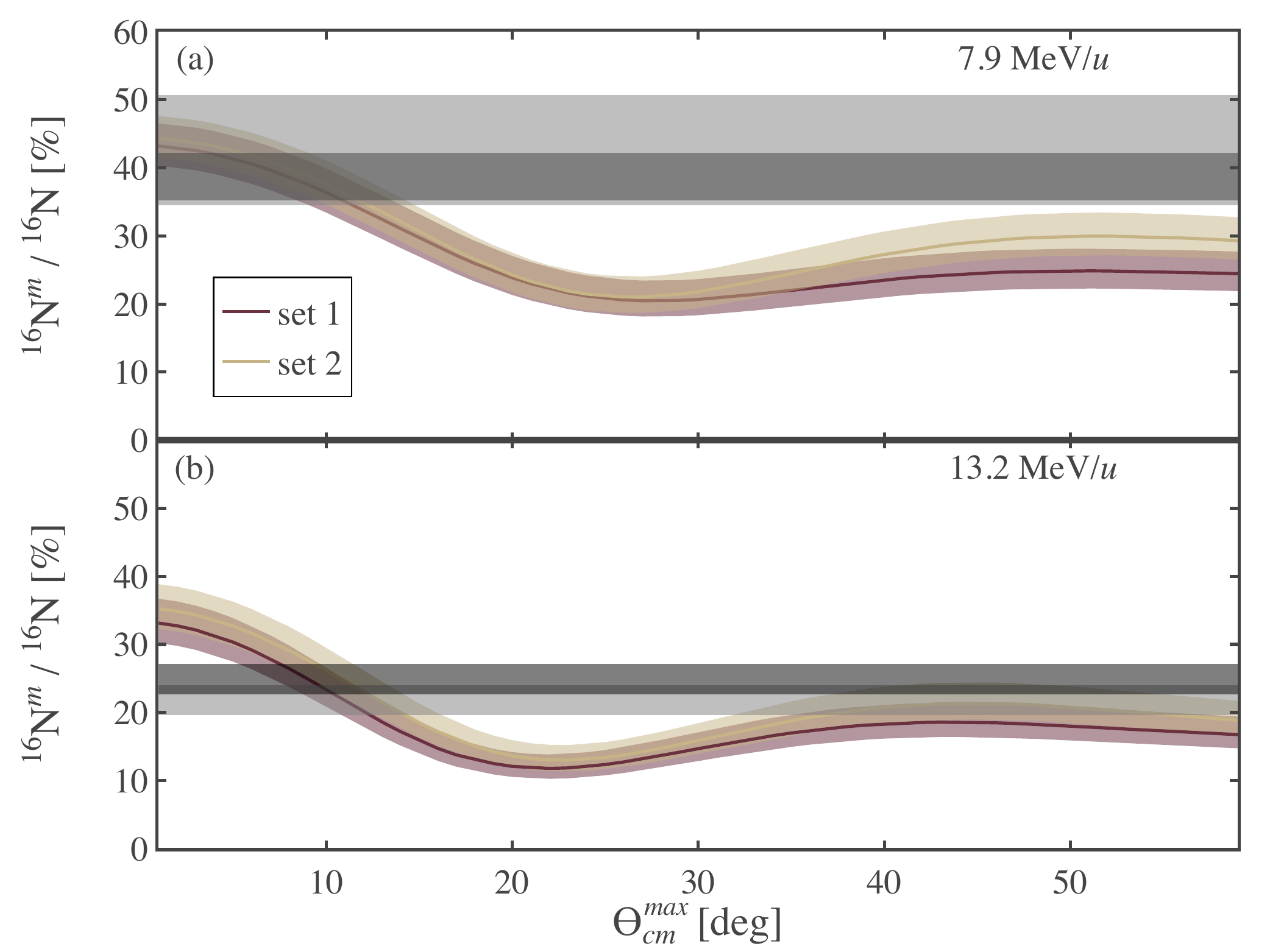}
	\caption{The calculated $^{16}$N$^{m}$/$^{16}$N isomer percent as a function of the maximum integration angle, $0\leq\Theta_{cm}\leq\Theta_{cm}^{max}$, for reaction energies of 7.9~MeV/$u$ (a) and 13.2~MeV/$u$ (b). The lines, shaded areas, and colors for the calculations are the same as those described in Fig.~\ref{fig:results}. The measured isomer fractions are taken from Fig.~\ref{fig:results} and are represented by the horizontal  bands, with the 120-keV/1755-keV data in light grey and the 120-keV/2742-keV data in dark grey.
	}
	\label{fig:isocm}
\end{figure}

Calculation of the cross sections, $\sigma$, up to a specified maximum center-of-mass angle, $\Theta_{cm}^{max}$, were carried out for each of the bound states populated in $^{16}$N using the DWBA approach outlined in Section~\ref{sec:meth}.
Each $\sigma$ value was multiplied by its corresponding relative $S_{\ell}$ value belonging to the $^{15}$N ground state overlap to each level in $^{16}$N.
Empirical determination of the relative $S_{\ell}$ suggest that the $S_{2}$ values may be reduced by as much as 30\% relative to $S_{0}$~\cite{ref:Warburton1957,ref:Bohne1972,ref:Bardayan2008}.
The range for this ratio, $S_{2}/S_{0} = 0.7 - 1.0$, was accounted for in the calculations of the isomer fraction and defines the width of the bands on the calculations shown in Figs.~\ref{fig:results}(a) and (b).
The $^{16}$N $J^{\pi}=3^-$ and 1$^-$ bound excited states are known to have short lifetimes (T$_{1/2}\lesssim 100$~ps) and decay promptly,  relative to the beam time-of-flight, after being populated.
The uncertainties in the $\gamma$-ray branching ratios from the $1^-$ state are on the order of a few percent, small compared to the variation in the spectroscopic overlaps.
Finally, the decay of the isomer throughout the flight time from the production target to the experimental station ($\approx30$~m) was accounted for.
A decreasing amount, from 12\% to 5\% of the directly populated isomer yield, decays into the ground state during the flight time over the reaction energy range of 3 -- 16~MeV/$u$. 
The calculated isomer fraction as a function of the reaction energy is shown in Fig.~\ref{fig:results} and as a function the maximum integration angle in Fig.~\ref{fig:isocm}.

\subsection{the dependencies of the isomer fraction \& composite rate on $\Theta_{cm}^{max}$}
A key factor in the DWBA description of the observed isomer ratio data was the proper upper-limit on the center-of-mass angle integration range ($0^{\circ} \leq \Theta_{cm} \leq \Theta_{cm}^{max}$). The calculated dependence of $^{16}$N$^{m}$/$^{16}$N on $\Theta_{cm}^{max}$ is illustrated in Fig.~\ref{fig:isocm}. A value of $\Theta_{cm}^{max}\approx10^{\circ}$ ($\Theta_{cm}^{max}=9^{\circ}$ and $11^{\circ}$ for sets 1 and 2, respectively) was determined to best reproduce the reaction-energy dependence of the measured data (Fig.~\ref{fig:results}). While $\Theta_{cm}^{max}$ values ranging between a few degrees up to $\approx15^{\circ}$ were also consistent with the measured isomer fraction data. At both reaction energies and for both parameter sets, the calculated isomer fractions evolve in similar fashion and give extreme values that span a factor of 2. Fig.~\ref{fig:isocm} solidified the need to include $\Theta_{cm}^{max}$ as a key degree of freedom when attempting to describe the isomer fraction.

The underlying cause for the behaviour of the $^{16}$N$^{m}$/$^{16}$N fraction as a function of $\Theta_{cm}^{max}$ is once again the differing $d\sigma/d\Omega$ for the $\ell=0$ and $2$ neutron transfers (Fig.~\ref{fig:ad}).
When $\Theta_{cm}^{max}$ is reduced, and the integration window closes, more relative yield is found in the $\ell=0$ cross sections. 
Most of the $\ell=0$ $\sigma$ feeds into the $0^-$ isomeric state, while 100\% of the $\ell=2$ strength feeds into the 2$^-$ ground state, hence, the result is an increase in $^{16}$N$^m$/$^{16}$N.
As previously mentioned, the mass (binding energy) and beam momenta difference between a $^{16}$N$^{m}$ ion and a $^{16}$N$^{g}$ ion is small with no discernible difference in their transport.
However, the size of the angular acceptance of the $^{16}$N ions beam \emph{does} impact the $^{16}$N$^{m}$/$^{16}$N ratio.

Based upon the known angular and momentum acceptances of RAISOR, $\Theta_{cm}^{max}$ should be around 40-50$^{\circ}$ up to its focal plane.
 However, the elements in the transport beam line from the RAISOR focal plane to the experimental station have created an effective acceptance less than this. The limitation in the beam transport line is likely in the angular acceptance as opposed to the momentum acceptance due to a number of known, and constraining, apertures, e.g., beam-profile scanners and the access points of the re-bunching cavity cryostats.

\begin{figure}[t]
	\centering
    \includegraphics[width=0.48\textwidth]{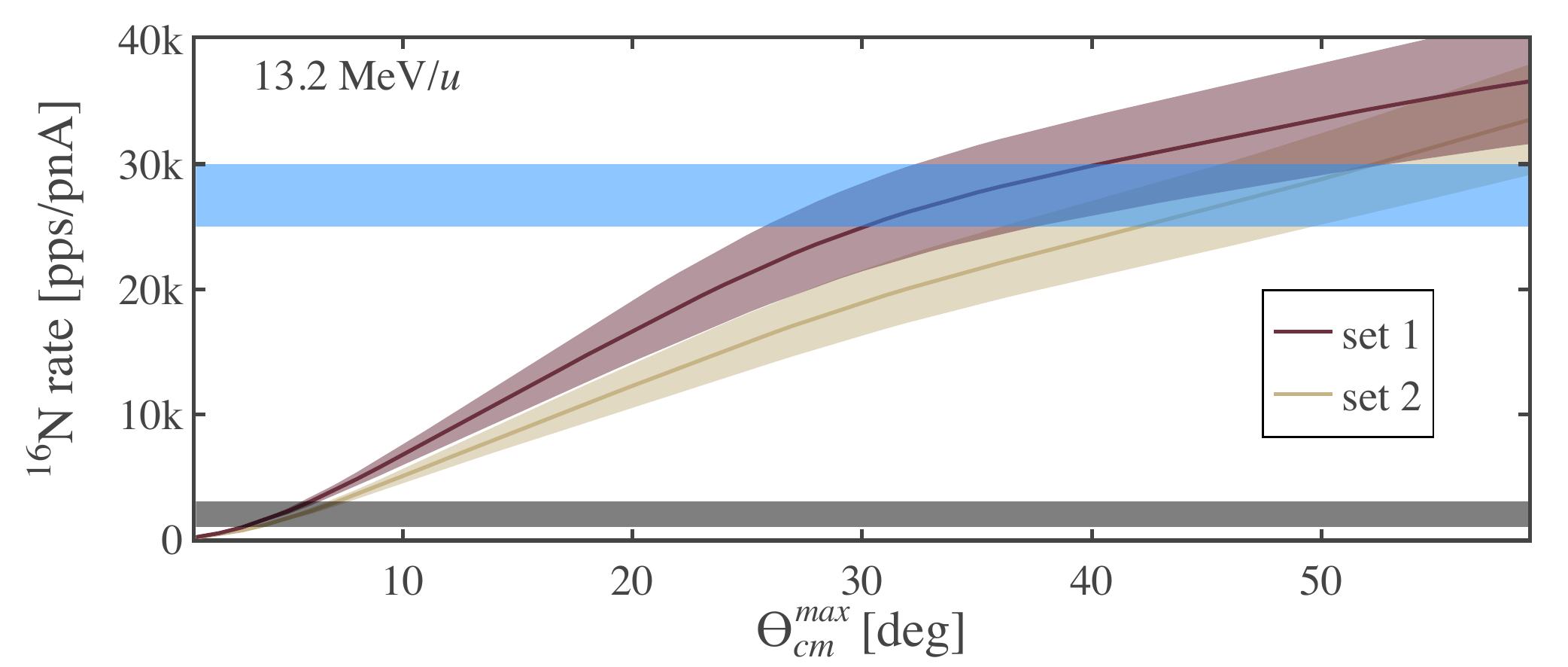}
	\caption{The calculated $^{16}$N composite beam rate as a function of the maximum integration angle, $0\leq\Theta_{cm}\leq\Theta_{cm}^{max}$, at a reaction energy of 13.2~MeV/$u$. The lines, shaded areas, and colors for the calculations are the same as those described in Fig.~\ref{fig:results}. There is an additional 20\% uncertainty on the calculated bands that is not shown due to assumptions in the quenching factor on the absolute $S_{\ell}$ values and sensitivities in the parameters used by the DWBA calculations. The range of the measured $^{16}$N rate at the RAISOR focal plane is given by the blue band. The measured $^{16}$N rate at the experimental station is given by the dark grey band.
	}
	\label{fig:ratecm}
\end{figure}

Another exploration into the available data is shown in Fig.~\ref{fig:ratecm} where $^{16}$N$^{m}$/$^{16}$N is given as a function of the $^{16}$N composite beam rate. The measured rates for the 13.2~MeV/$u$ beam energy are shown by the two bands, one taken at the RAISOR focal plane (blue band) and the other at the experimental target station where the isomer content was measured (dark-grey band). The DWBA approach also provided a calculation of the absolute rates, the results of which are shown by the garnet and gold bands in Fig.~\ref{fig:ratecm} for the two optical model parameter sets. Such calculations are reliable for single-neutron reactions, particularly when the relative $S_{\ell}$ values have been extracted from measurement. The calculation of the total beam rate included all of the components used in the isomer fraction estimates. The relative values of $S_{\ell}$ were multiplied by an empirically established quenching factor of 0.55~\cite{ref:Aum21} to obtain their absolute values. From the target dimensions, temperature, and pressure, $\approx 4.4\times 10^{20}$ deuterium target atoms / cm$^2$ were assumed in the calculations. There is an additional systematic uncertainty on the order of 20\% on the rate calculations which is not shown in Fig.~\ref{fig:ratecm}. This is due to uncertainties in the values of the quenching factor and various optical model parameters in the DWBA approach. Not surprisingly, as $\Theta_{cm}^{max}$ increases in Fig.~\ref{fig:ratecm} $\sigma$ follows and the calculated rate increases.

The data collected at the experimental station (the dark-grey band in Fig.~\ref{fig:ratecm}) overlaps with the rate calculations for small integration ranges, $\Theta_{cm}^{max}\lesssim 8^{\circ}$. 
The composite $^{16}$N beam rate measured at the RAISOR focal plane for the 13.2~MeV/$u$ reaction energy (blue band), however, overlaps with $\Theta_{cm}^{max}\gtrsim25^{\circ}$ for DWBA parameter set 1, and $\Theta_{cm}^{max}\gtrsim35^{\circ}$ for set 2. Both are in the vicinity of the limits from RAISOR alone ($\Theta_{cm}^{max}\approx45-50^{\circ}$) imposed by the expected momentum and angular acceptance. The $^{16}$N rate data is consistent with the conclusions drawn from the comparisons between the calculations and data in Figs.~\ref{fig:results} and~\ref{fig:isocm}. Namely, that components in the transport beam line have led to an effectively reduced angular acceptance for the $^{16}$N beam relative to that expected by RAISOR alone. The impact, then, is a reduced $\Theta_{cm}^{max}$, which influences both the isomer fraction ($^{16}$N$^{m}$/$^{16}$N increased) and the total $^{16}$N rate (reduced) when measured at the experimental station.

\section{Summary \& Conclusions}
A beam of $^{16}$N was produced via the single-neutron adding reaction $^{15}$N($d$,$p$)$^{16}$N in inverse kinematics at the upgraded ATLAS in-flight system. The $^{16}$N$^{m}$/$^{16}$N fraction for the $J^{\pi}=0^-$ isomeric state was measured to be $\approx40$\% and $\approx24$\% at reaction energies of 7.9(3)~MeV/$u$ and 13.2(2)~MeV/$u$, respectively. An Al stopping foil and HPGe Clover detector were used in concert to stop and count the number of $\gamma$ rays depopulating the isomeric state as well as the number of $\gamma$ rays following the $\beta$ decay of the $^{16}$N ground state. This took place at an experimental station tens of meters downstream of the focal plane of the magnetic separator (RAISOR). The composite rate of the $^{16}$N beam was measured at the focal plane of the RAISOR separator for the 13.2~MeV/$u$ reaction energy and at the experimental station for both reaction energies.

Estimates of the isomer fraction and the total $^{16}$N beam rates were made using DWBA calculations of the reaction cross sections and the known spectroscopic information on $^{16}$N. The measured data retrieved from the experimental station was reproduced only when a maximum center-of-mass integration angle of $\Theta_{cm}^{max}\lesssim10^{\circ}$ was applied to the differential cross sections. However, the measured rates at the RAISOR focal plane were consistent with larger $\Theta_{cm}^{max}$, which would yield differing isomer fractions. Therefore, while the transport efficiency is the same for both $^{16}$N$^m$ and $^{16}$N$^g$ ions - their total momenta are essentially the same - the differing angular dependencies involved in their production gives rise to a sensitivity in the $^{16}$N$^{m}$/$^{16}$N fraction as a function of angular acceptance.

To conclude, the present work demonstrates how changes in the angular acceptance provide a degree of freedom, in addition to the reaction energy, in which to control the isomer content of a beam produced via single-nucleon transfer reactions. Various hardware modifications to the ATLAS in-flight facility are currently being discussed to leverage this additional degree of freedom for cases where, for example, the energy requirements of the secondary beam are fixed. For instance, an adjustable aperture could be inserted at the exit of the final RAISOR quadrupole, which is equivalent to the entrance quadrupole aperture considering RAISOR is a magnetic achromat, to control angular acceptance. In parallel, the downstream beam line from RAISOR will be explored to define and remove the limiting apertures in order to approach a similar acceptance to that of RAISOR. Finally, a measurement of the $^{16}$N$^{m}$/$^{16}$N fraction at the RAISOR focal plane would provide a confirming test of the isomer fraction dependency on the angular acceptance.

\section*{Acknowledgements}
The authors would like to acknowledge the efforts of the ATLAS
Facility’s operations team, P.~Copp, B.~Mustapha, and C.~Dickerson towards this work. This research used resources of Argonne National Laboratory’s ATLAS facility, which is a Department of Energy Office of Science User Facility. This material is based upon work supported by the U.S. Department of Energy, Office of Science, Office of Nuclear Physics, under Contract Nos. DE-AC02-06CH11357 (ANL), DE-AC05-00OR22725 (ORNL), DE-SC0020451 (NSCL), and Grant No. DE-FG02-96ER40978 (LSU). This work was also supported by the Japanese Hirose International Scholarship Foundation.

\bibliographystyle{elsarticle-num} 
\bibliography{raisorbib,references}
\end{document}